\documentclass[prl,twocolumn,aps,showpacs]{revtex4}
\usepackage{epsfig}
\usepackage{bm}
\begin{document}
\date{\today}
\title{Effects of density imbalance on the BCS-BEC crossover
in semiconductor electron-hole bilayers}
\author{P. Pieri,$^1$  D. Neilson,$^{1,2}$ and G.C. Strinati$^1$}
\affiliation{\mbox{$^1$Dipartimento di Fisica, Universit\`{a} di Camerino, I-62032 Camerino, Italy}\\
\mbox{$^2$School of Physics, University of New South Wales, Sydney 2052, Australia}}
\begin{abstract}
We study the occurrence of excitonic superfluidity in electron-hole
bilayers at zero temperature. 
We not only identify the crossover in the phase diagram from the BCS limit of overlapping
pairs to the BEC limit of non-overlapping tightly-bound pairs but
also, by varying the electron and hole
densities \emph{independently}, we can analyze a number of phases that occur mainly in the
crossover region. With different electron
and hole effective masses, the phase diagram is asymmetric with
respect to excess electron or hole densities. We propose as the
criterion for the onset of superfluidity, the jump of the electron
and hole chemical potentials when their densities cross.
\end{abstract}
\maketitle

There is a growing interest in the physics of the BCS-BEC crossover,
owing to the recent experimental advances with trapped Fermi atoms.
With the use of Fano-Feshbach resonances, this crossover has been
observed with fermionic $^{6}$Li and $^{40}$K
atoms~\cite{exp_crossover} which become composite bosons in the BEC
limit. Although trapped atoms represent an ideal testing ground for
a fundamental understanding of the BCS-BEC crossover, technological
applications exploiting the occurrence of condensates will most
probably rely on semiconductor systems. In these systems, excitons
made up of electrons and holes play the role of composite bosons.

Excitonic systems were, in fact, the first to be considered for the
BCS-BEC crossover. In bulk materials, original work on exciton
condensation was done by Keldysh and co-workers
\cite{Keldysh-65-68}. Extension of this to the BCS-BEC crossover was
proposed by Nozi\`{e}res and Comte \cite{Nozieres-82}. However, in
bulk materials fast electron-hole recombination hinders the
detection of BEC. Proposals have accordingly been made to condense
excitons with spatially separated electrons and holes
\cite{Shevchenko-1976,Lozovik-1975-76}. Particularly promising are
bilayer quantum-well systems separated by a distance $d$, with
conduction-band electrons in one well and valence-band holes in the
adjacent well \cite{david,Rice-1995,gortel,gaetano}. Recently,
developments for the detection of excitonic BEC in coupled quantum
wells has been reported
\cite{exp-BEC-excitons,half-full-empty,MacDonald-2004}.
Technological advances with electron-hole bilayers
\cite{vonKlitzing-2002} make it now possible to contact
\emph{separately} the layers of electrons and holes in GaAs,
separated by an AlGaAs barrier of thickness $d$ as small as 15 nm.

In this paper, we consider the BCS-BEC crossover in electron-hole
bilayers when the densities of the electrons and holes are varied
independently of each other. The effect of the density imbalance
resembles that of a magnetic field in a superconductor (disregarding
orbital effects) first considered by Sarma~\cite{Sarma}. This
analogy was also noted in Ref.~\cite{density-imbalance-nuclei},
where the influence of an isospin asymmetry in nuclear matter was
considered for the BCS-BEC crossover. The specific system we shall
consider as an example is GaAs-AlGaAs.

Recent experiments with population imbalance in ultracold trapped
Fermi atoms~\cite{Ketterle-Hulet-2006} have stimulated a
considerable amount of theoretical work on two-component Fermi
systems with density
imbalance~\cite{theoretical-density-imbalance-hom,theoretical-density-imbalance-trap}.
Emphasis has been placed on the possible occurrence of exotic phases
in addition to the ordinary BCS
pairing~\cite{theoretical-density-imbalance-hom,Wilczek,FFLO&Co}.
However, in these systems the presence of a trap and their charge
neutrality inhibit the occurrence of the exotic phases, and so far
only phase separation between a superfluid core with equally matched
populations and an outer normal phase has been detected.

Electron-hole bilayers may offer a better opportunity of observing
such exotic phases because the Coulomb repulsion within each layer acts to suppress phase
separation~\cite{coulomb}. In addition, we find that the different
electron and hole effective masses in GaAs, $m_{\mathrm{e}}$ and
$m_{\mathrm{h}}$, and also the non-local nature of the electron-hole
attraction both favor the occurrence of exotic phases. These include
the Sarma phase with one or two Fermi surfaces and finite population
imbalance, and the Fulde-Ferrel-Larkin-Ovchinnikov (FFLO)
phase~\cite{FFLO}, as well as the ordinary BCS pairing with equal
populations. The relative extension of these phases is quite
asymmetric between an imbalance with more holes and an imbalance
with more electrons. We also find that in the superfluid phase the
separate electron and hole chemical potentials display a \emph{jump}
when reversing the population imbalance from more electrons to more
holes, while in the normal phase no jump occurs. Detection of this
jump could thus serve to identify superfluid character in a system.



The electron-hole bilayer system is described by the grand-canonical Hamiltonian:
\begin{eqnarray}
&&K=\sum_{\mathbf{k},\sigma} \xi_{\mathbf{k} \sigma}
c^{\dagger}_{\mathbf{k} \sigma} c_{\mathbf{k} \sigma} +
\frac{1}{2\Omega}
\sum_{\mathbf{k},\mathbf{k'},\mathbf{q},\sigma,\sigma'} V^{\sigma \sigma'}_{\mathbf{k}-\mathbf{k'}}\nonumber\\
&&\times \,
 c^{\dagger}_{\mathbf{k}+\frac{\mathbf{q}}{2} \sigma}
c^{\dagger}_{-\mathbf{k}+\frac{\mathbf{q}}{2} \sigma'}
c_{-\mathbf{k'}+\frac{\mathbf{q}}{2} \sigma'}
c_{\mathbf{k'}+\frac{\mathbf{q}}{2} \sigma}\, .
\label{Grand-canonical-Hamiltonian}
\end{eqnarray}
Here, $\mathbf{k}$, $\mathbf{k'}$, and $\mathbf{q}$ are
two-dimensional wave vectors in the layers, $\Omega$ is the
quantization volume (surface area), $c^{\dagger}_{\mathbf{k}
\sigma}$ ($c_{\mathbf{k} \sigma}$) are the creation (destruction)
operators for electrons ($\mathrm{e}$) and holes ($\mathrm{h}$)
distinguished by $\sigma=(\mathrm{e},\mathrm{h})$, the
$\xi_{\mathbf{k} \sigma}= \epsilon_{\mathbf{k} \sigma} -
\mu_{\sigma}$ are the band dispersions with chemical potentials
$\mu_{\sigma}$ for electrons and holes, $\epsilon_{\mathbf{k}
\mathrm{e}}=\mathbf{k}^{2}/(2 m_{\mathrm{e}}) + E_{\mathrm{g}}$ and
$\epsilon_{\mathrm{h}}(\mathbf{k})=\mathbf{k}^{2}/(2
m_{\mathrm{h}})$. The semiconductor band gap $E_{\mathrm{g}}$ can be 
reabsorbed in the electron chemical potential.
Explicit spin quantum numbers are omitted.

We have carried out our calculations in the zero-temperature limit,
where a mean-field description of the BCS-BEC crossover is
appropriate even in two dimensions~\cite{Nozieres-Pistolesi-99}. We
use an unscreened electron-hole attractive potential as in
Ref.~\cite{Rice-1995},
$V^{\mathrm{e}\mathrm{h}}_{\mathrm{k}}= - 2 \pi e^{2} \exp{
(-kd)}/(k \varepsilon)$, where $k=|\mathbf{k}|$, $e$ is the electron
charge, and $\varepsilon$ the background dielectric constant. The
gap in the superfluid phase makes screening less
effective, so introducing screening should only cause small
quantitative changes.

In the present calculation we neglect the intra-layer Coulomb
repulsions $V^{\mathrm{e}\mathrm{e}}$ and
$V^{\mathrm{h}\mathrm{h}}$. In the absence of density imbalance,
$V^{\mathrm{e}\mathrm{e}}$ and $V^{\mathrm{h}\mathrm{h}}$ can be
readily included in a mean-field treatment~\cite{Rice-1995}, but
when the densities are imbalanced, including
$V^{\mathrm{e}\mathrm{e}}$ and $V^{\mathrm{h}\mathrm{h}}$ in the
mean field requires a detailed knowledge of how overall charge
neutrality is attained. This is in order to avoid divergence of the
Hartree term. This depends on the specific engineering configuration
of the device, and therefore in the interests of generality we drop
$V^{\mathrm{e}\mathrm{e}}$ and $V^{\mathrm{h}\mathrm{h}}$ here. When
the densities are equal we can compare with Ref.~\cite{Rice-1995},
and we have verified that omitting
$V^{\mathrm{e}\mathrm{e}}$ and $V^{\mathrm{h}\mathrm{h}}$ reduces
the size of the gap by an amount no larger than 30 \%. As a separate
issue, and as we have already noted, intra-layer repulsion should
stabilize the system against phase separation. For this reason, we
exclude the possibility of phase separation from our discussion of
the phase diagram.


The relevant mean-field equations to be solved for the variables $\mu_{\mathrm{e}}$, $\mu_{\mathrm{h}}$, and the (s-wave) gap function $\Delta_{\mathbf{k}}$ are:
\begin{eqnarray}
\Delta_{\mathbf{k}} &=& - \frac{1}{\Omega} \sum_{\mathbf{k'}} V^{\mathrm{e}\mathrm{h}}_{\mathbf{k}-\mathbf{k'}}
\frac{\Delta_{\mathbf{k'}}}{2 E_{\mathbf{k'}}}
[1 - f(E^{+}_{\mathbf{k'}}) - f(E^{-}_{\mathbf{k}'})], \label{Delta-eqn}\\
n_{\mathrm{e}} &=& \frac{1}{\Omega} \sum_{\mathbf{k}}
\left[ u_{\mathbf{k}}^{2} f(E^{+}_{\mathbf{k}}) + v_{\mathbf{k}}^{2} ( 1 - f(E^{-}_{\mathbf{k}})) \right], \label{n-electron}\\
n_{\mathrm{h}} &=& \frac{1}{\Omega} \sum_{\mathbf{k}}
\left[ u_{\mathbf{k}}^{2} f(E^{-}_{\mathbf{k}}) + v_{\mathbf{k}}^{2} ( 1 - f(E^{+}_{\mathbf{k}})) \right], \label{n-hole}
\end{eqnarray}
where $f(E) = \Theta(-E)$ is the Fermi function at zero temperature,
$E_{\mathbf{k}} = \sqrt{\xi_{\mathbf{k}}^2 +
\Delta_{\mathbf{k}}^{2}}$ with $\xi_{\mathbf{k}} =
(\xi^{\mathrm{e}}_{\mathbf{k}} + \xi^{\mathrm{h}}_{\mathbf{k}})/2$,
and $E^{\pm}_{\mathbf{k}} = E_{\mathbf{k}} \pm \delta
\xi_{\mathbf{k}}$ with $\delta \xi_{\mathbf{k}} =
(\xi^{\mathrm{e}}_{\mathbf{k}} - \xi^{\mathrm{h}}_{\mathbf{k}})/2$.
In addition, $v_{\mathbf{k}}^{2} = 1 - u_{\mathbf{k}}^{2} = ( 1 -
\xi_{\mathbf{k}}/E_{\mathbf{k}})/2$. Throughout, we express lengths
in units of the effective Bohr radius $a_{0}^{*}$, and energies in
units of the effective Rydberg Ry$^{*}$ \cite{footnote-1}. In two
dimensions the average interparticle spacing is $r_{s}=[\pi
(n_{\mathrm{e}}+n_{\mathrm{h}})/2]^{-1/2}$.

\begin{figure}
\begin{center}
\epsfxsize=7.05cm
\epsfbox{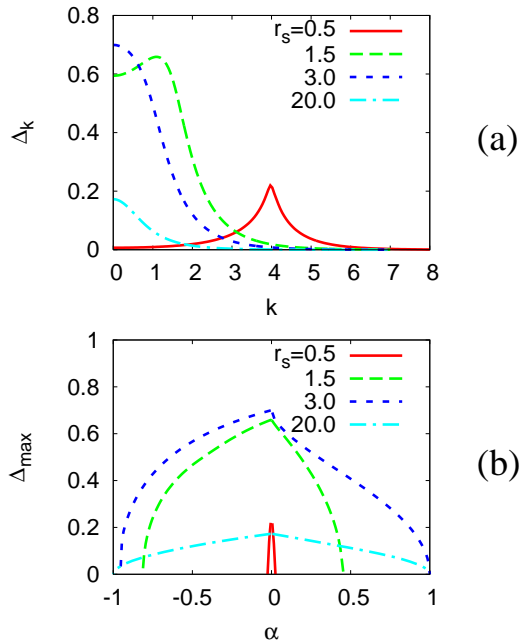}
\vspace{0.5cm}
\caption{(a) Wave-vector dependence of the gap function for $\alpha=0$,
$d=1$, and several values of $r_{s}$; (b) Maximum value $\Delta_{\rm{max}} =
\mathrm{\rm{max}} \{ \Delta_{\mathbf{k}} \}$ vs $\alpha$ for $d=1$ and several
values of $r_{s}$.}
\end{center}
\end{figure}

Figure 1(a) shows the wave-vector dependence of
$\Delta_{\mathbf{k}}$ for equal densities, layer separation $d=1$,
and several values of $r_s$. These values of $r_{s}$ 
and a layer separation of $d=1$ are already experimentally
attainable. The wave-vector position $k_{\rm{max}}$ of the peak in
the gap function is seen to evolve from a finite value in the BCS
regime (small $r_s$) toward zero in the BEC regime (large $r_s$),
while the corresponding value $\Delta_{\rm{max}}$ of the gap
function attains its maximum for intermediate values of $r_{s}$.
This is a generic feature of the density-induced BCS-BEC
crossover~\cite{Natascia-99}.

We have verified for several values of the distance $1\le d\le 4$
that the optimal value of $r_s$ at which $\Delta_{\rm{max}}$
attains its largest value is located close to the value of $r_s$
where the average chemical potential $\mu = ( \mu_{\mathrm{e}} +
\mu_{\mathrm{h}} )/2$ crosses zero. For $d=1$ this occurs near
$r_s=3$. This finding is in line with results for the BCS-BEC
crossover with a contact potential in three dimensions, namely, that
superfluid properties are more robust in the region located between
the Fano-Feshbach resonance and the vanishing of the chemical
potential \cite{Natascia-03}. The narrow region lying between the
optimal value of $r_s$ and the value corresponding to $\mu=0$ thus
identifies the middle of the crossover region between the BCS and
BEC regimes for the electron-hole bilayer.

The effect of the density imbalance $\alpha \equiv (n_{\mathrm{e}} -
n_{\mathrm{h}})/(n_{\mathrm{e}} + n_{\mathrm{h}})$ on
$\Delta_{\rm{max}}$ is shown in Fig.~1(b). We see that the density
imbalance acts to reduce the magnitude of the energy gap and that it
has different effects on the two sides of the crossover. In the BCS
regime, the mismatch of the Fermi surfaces for electrons and holes
strongly affects the superfluid properties of the system. The
superfluid properties are lost when the Fermi energies mismatch
becomes large compared with the value of $\Delta_{\rm{max}}$ for
equal densities. For $r_s$ small this occurs for a small value of
$\alpha$. In the BEC regime, the superfluid properties are less
sensitive to density imbalance, whose main effect is then to reduce
the number of electron-hole pairs. 

\begin{figure}
\begin{center}
\epsfxsize=6.2cm 
\epsfbox{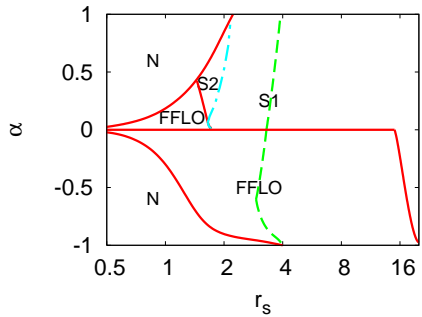} \vspace{0.5cm} 
\caption{Zero-temperature phase diagram 
for $d=1$, showing the stability domains of the different phases: Normal
(N), Sarma (S1 and S2), and FFLO. The dashed line corresponds to the
curve $\mu=0$, and the dash-dotted line separates the S1 and S2 phase.
Note the logarithmic scale for $r_s$.}
\end{center}
\end{figure}

Figure 2 shows the zero-temperature phase diagram for $d=1$. We can
identify various phases using $\Delta_{\mathbf{k}}$, determined from
Eq.~(2), and the superfluid (mass) density $\rho_s$. Within mean-field
theory and in the zero-temperature limit, $\rho_{s}$ is given by:
\begin{equation}
\rho_{s} \, = \, m_{\mathrm{e}} n_{\mathrm{e}} \, + \,
m_{\mathrm{h}} n_{\mathrm{h}} \, - \, \frac{1}{4 \pi} \,
\sum_{j,\lambda} \, \frac{(k_{j}^{\lambda})^{3}}{\left|
\frac{dE^{\lambda}_k}{dk} \right|_{k=k_{j}^{\lambda}}} \, .
\label{superfluid-density}
\end{equation}
Here, $k_{j}^{\lambda}$ is the $j$-th zero of $E^{\lambda}_k=0$ with
$\lambda=(+,-)$. [For positive (negative) density imbalance only
$E^{+}_k$ ($E^{-}_k$) has zeros, while no zero occurs for $\alpha =
0$.]

The normal phase (N) corresponds to the trivial solution
$\Delta_{\mathbf{k}}=0$. The Sarma phases corresponds to
nonvanishing $\Delta_{\mathbf{k}}$ when $\alpha \ne 0$ and positive
superfluid density $\rho_s$. The S1 and S2 denote the Sarma phases
for one and two Fermi surfaces, respectively. There will be one zero
of $E^{\lambda}_k$ ($j=1$) for the S1 phase (one Fermi surface), and
two zeros ($j=1,2$) for the S2 phase (two Fermi surfaces).
Sometimes, the Sarma S2 phase is called the ``breached-pair'' phase
after Ref.~\cite{Wilczek}. A negative value of $\rho_s$ in Eq.~(5)
indicates that the Sarma phase is unstable toward a phase with a
spontaneously generated superfluid current, which we
associate~\cite{Yip-03,cinesi-06} with the FFLO phase. We have
verified that the Sarma phase, whenever it exists, is always lower
in energy with respect to the normal phase. (We recall in this
respect that our calculation is at fixed density imbalance,
while the original Sarma calculation was at fixed chemical
potentials.)


The most prominent feature of Fig.~2 is the marked dependence of the
phase diagram on the sign of $\alpha$~\cite{SadeMelo-06}. In
particular, while the boundary of the normal phase does not depend
appreciably on the sign of $\alpha$, the region of stability of the
Sarma phase with respect to the FFLO phase depends dramatically on
the sign. For $\alpha < 0$, the phase diagram is dominated by the
FFLO phase, with the Sarma (S1) phase being confined to the extreme
BEC region, while for $\alpha > 0$, the FFLO phase is compressed
into the region of small $r_s$.

Such an asymmetry can be understood in terms of the relevant
dispersion $E^{+}_{\mathbf{k}}$ ($E^{-}_{\mathbf{k}}$) for positive
(negative) $\alpha$. Due to the mass difference, the term $2 \delta
\xi_{\mathbf{k}}$ which is added (subtracted) to $E_k$ makes the
dispersion $E^{+}_{\mathbf{k}}$ ($E^{-}_{\mathbf{k}}$) steeper
(flatter) with respect to the case of equal masses. As it is clear
from the last term in Eq.~(\ref{superfluid-density}), a flatter
dispersion will make the superfluid density more negative, thus
replacing the Sarma phase with the FFLO phase. The opposite occurs
for positive $\alpha$, when the relevant dispersion becomes
steeper.

In this part of the phase diagram there is some room even for the S2
Sarma phase. So far, this phase was found in the literature to be
very fragile, being invariably replaced by phase separation or by
the development of a FFLO phase~\cite{Bedaque-03,Yip-03,Wilczek}.
Here, the concurrence of several favorable factors, the mass
difference, the wave-vector dependence of the gap, and the
intralayer Coulomb repulsion, all serve to stabilize the S2 Sarma
phase in an appreciable region of the phase diagram.

We note that the largest number of phases occurs for intermediate
values of $r_s$ (say, $r_s=1.5 - 3$), corresponding to the smooth
crossover region between the BCS and BEC limit for $\alpha=0$. For
non-zero $\alpha$, several transition lines can be crossed in this
region by varying $\alpha$ or $r_s$. This is thus the most fertile
region to be explored experimentally, also because we recall in the
same region $\Delta_{\rm{max}}$ is largest and the superfluidity
most robust.

In this respect, the behavior of the separate chemical potentials
$\mu_{\mathrm{e}}$ and $\mu_{\mathrm{h}}$ vs $\alpha$ can serve to
reveal the appearance of a finite value for the gap $\Delta$. This
is because $\mu_{\mathrm{e}}$ and $\mu_{\mathrm{h}}$ must show a
jump across $\alpha=0$ in order to sustain a finite density
imbalance in the superfluid phase. This behavior is shown in Fig.~3
for $d=1$ and $r_{s}=3$, when $\mu_{\mathrm{e}}(\alpha=0^{+}) \, -
\, \mu_{\mathrm{e}}(\alpha=0^{-}) \, \simeq \, 2 \,
\Delta_{\rm{max}}$. The possible instability toward the FFLO phase
we discussed above should not affect the occurrence of this jump.
Physical quantities obtained for the Sarma and the FFLO phases
should, in fact, merge continuously when $\alpha$ approaches zero,
owing to the corresponding vanishing of the modulation wave vector
associated with the FFLO phase \cite{cinesi-06}.

It is useful to discuss the sensitivity of our results to the
parameters used in the calculations. Increasing the distance $d$
will shift the location of the intermediate region where most phases
are seen, to larger values of $r_s$ (for $d=2$, e.g., the shift is
about a 30\%), but it does not alter the shape of the phase diagram.
Reducing the mass difference has the effect of contracting the FFLO
phase for negative $\alpha$ and expanding it for positive $\alpha$.
This is as expected from the arguments discussed above. However, even
for equal masses the phase diagram changes only quantitatively,
allowing even in this case some space for the S2 Sarma phase. Finite
temperature assists in stabilizing the Sarma phases with respect to
the FFLO: the superfluid density is, in fact, quite sensitive to
temperature in the presence of density imbalance. We have found,
however, that temperatures below $1$ K do not appreciably alter the
phase diagram in the most interesting intermediate $r_s$ region.
Finally, a reduction of the FFLO phase due to the intra-layer
Coulomb repulsion may occur but this reduction should be limited to
the region of large $r_{s}$, where the spatial modulation of the gap
parameter is expected to be accompanied by a density modulation.

\begin{figure}
\begin{center}
\epsfxsize=6.2cm
\epsfbox{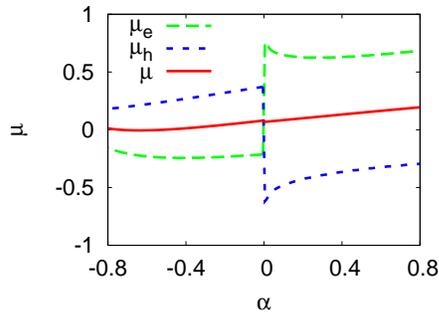}
\caption{Zero-temperature electron
($\mu_{\mathrm{e}}$) and hole ($\mu_{\mathrm{h}}$) chemical potentials vs
$\alpha$ for $d=1$ and $r_{s}=3$.
The average chemical potential $\mu$ is also shown.}
\end{center}
\end{figure}

In conclusion, we have shown that electron-hole bilayers are
promising candidates for revealing a variety of superfluid phases
while traversing the BCS-BEC crossover. Exotic phases such as the
FFLO and the S2 Sarma phases which have been so far elusive to
experimental detection, should be notably robust in electron-hole
bilayer systems.

\acknowledgments
This work was partially supported by the Italian
MIUR with contract Cofin-2005 ``Ultracold Fermi Gases and Optical
Lattices''.




\begin{thebibliography}{99}


\bibitem{exp_crossover} M. Bartenstein {\em et al.}, Phys. Rev. Lett. {\bf 92}, 120401 (2004); C.A. Regal, M. Greiner, and D.S. Jin, Phys. Rev. Lett.
                         {\bf 92}, 040403 (2004); M.W. Zwierlein {\em et al.}, Phys. Rev. Lett. {\bf 92} 120403 (2004).
\bibitem{Keldysh-65-68} L.V. Keldysh and Y.V. Kopaev, Sov. Phys. Solid State {\bf 6}, 2219 (1965);
                        L.V. Keldysh and A.N. Kozlov, Sov. Phys. JETP {\bf 27}, 521 (1968).

\bibitem{Nozieres-82} C. Comte and P. Nozi\`{e}res, J. Phys. (Paris) {\bf 43}, 1069 (1982);
                      P. Nozi\`{e}res and C. Comte, J. Phys. (Paris) {\bf 43}, 1083 (1982).


\bibitem{Shevchenko-1976} S.I. Shevchenko, Sov. J. Low. Temp. Phys. {\bf 2}, 251 (1976).

\bibitem{Lozovik-1975-76} Y.E. Lozovik and V.I. Yudson, JETP Lett. {\bf 22}, 274 (1975), and Sov. Phys. JETP {\bf 44}, 389 (1976).

\bibitem{david}
J. Szyma\'nski, L. \'Swierkowski, and D. Neilson, Phys. Rev. B {\bf 50}, 11002 (1994).


 \bibitem{Rice-1995} X. Zhu {\em et al.}, Phys. Rev. Lett. {\bf 74}, 1633 (1995).

\bibitem{gortel}
 Z.W. Gortel and L. \'{S}wierkowski, Surf. Science {\bf 361/362}, 146 (1996).

\bibitem{gaetano}
 S. De Palo, F. Rapisarda, and G. Senatore, Phys. Rev. Lett. {\bf 88}, 206401 (2002).

\bibitem{exp-BEC-excitons}
                           L.V. Butov, A.C. Gossard, and D.S. Chemia, Nature {\bf 418}, 751 (2002);
 A. T. Hammack {\em et al.}, Phys. Rev. Lett. {\bf 96}, 227402 (2006); S. Yiang, {\em et al.}, cond-mat/0606683.

\bibitem{half-full-empty}
 M. Kellogg {\em et al.}, Phys. Rev. Lett. {\bf 93}, 036801 (2004);
                          E. Tutuc, M. Shayegan, and D.A. Huse, Phys. Rev. Lett. {\bf 93}, 036802 (2004);
J.P. Eisenstein, Science {\bf 305}, 950 (2004).

\bibitem{MacDonald-2004} For a review, see
                        J.P. Eisenstein and A.H. MacDonald, Nature {\bf 432}, 691 (2004).


\bibitem{vonKlitzing-2002} M. Pohlt {\em et al.},
                           Applied Phys. Lett. {\bf 80}, 2105 (2002).

\bibitem{Sarma} G. Sarma, J. Phys. Chem. Solids {\bf 24}, 1029 (1963).

\bibitem{density-imbalance-nuclei} U. Lombardo {\em et al.\/}, Phys. Rev. C {\bf 64}, 064314 (2001).

\bibitem{Ketterle-Hulet-2006} 
M.W. Zwierlein {\em et al.}, Science {\bf 311}, 492 (2006);
G.B. Partridge {\em et al.}, Science {\bf 311}, 503 (2006);
Y. Shin {\em et al.}, Phys. Rev. Lett. {\bf 97}, 030401 (2006).


\bibitem{theoretical-density-imbalance-hom}
J. Carlson and S. Reddy, Phys. Rev. Lett. {\bf 95}, 060401 (2005); 
C.-H. Pao, Shin-Tza Wu, and S.-K. Yip, Phys. Rev. B {\bf 73}, 132506 (2006);
D. E. Sheehy and L. Radzihovsky, Phys. Rev. Lett. {\bf 96}, 060401 (2006).

\bibitem{theoretical-density-imbalance-trap}
 P. Pieri and G. C. Strinati, Phys. Rev. Lett. {\bf 96}, 150404 (2006); 
J. Kinnunen, L.M. Jensen, and P. T\"orm\"a, Phys. Rev. Lett. {\bf 96}, 110403 (2006); 
W. Yi and L. M. Duan, Phys. Rev. A {\bf 73}, 031604(R) (2006); 
F. Chevy, Phys. Rev. Lett. {\bf 96}, 130401 (2006);
T. N. De Silva and E. J. Mueller, Phys. Rev. A {\bf 73}, 051602(R) (2006); 
M. Haque and H. T. C. Stoof, Phys. Rev. A {\bf 74}, 011602(R) (2006);
J.-P. Martikainen, Phys. Rev. A {\bf 74}, 013602 (2006); 
C.-C. Chien {\em et al.}, Phys. Rev. A {\bf 74}, 021602(R) (2006).


\bibitem{Wilczek} M. McNeil Forbes {\em et al.}, Phys. Rev. Lett. {\bf 94}, 017001 (2005).

\bibitem{FFLO&Co} D. T. Son and M. A. Stephanov, Phys. Rev. A {\bf 74}, 013614 (2006); 
A. Bulgac, M. McNeil Forbes, and A. Schwenk, Phys. Rev. Lett. {\bf 97}, 020402 (2006);
M. Mannarelli, G. Nardulli, and M. Ruggieri, Phys. Rev. A {\bf 74}, 033606 (2006).

\bibitem{coulomb}
L.V. Butov {\em et al.}, JETP {\bf 92}, 260 (2001).

\bibitem{FFLO}
P. Fulde and R.A. Ferrell, Phys. Rev. {\bf 135}, A550 (1964);
A.I. Larkin and Yu.N. Ovchinnikov, Zh. Eksp. Teor. Fiz.
{\bf 47}, 1136 (1964) [Sov. Phys. JETP {\bf 20}, 762 (1965)].

\bibitem{Nozieres-Pistolesi-99} P. Nozi\`{e}res and F. Pistolesi, Eur. Phys. J. B {\bf 10}, 649 (1999).

\bibitem{footnote-1} For GaAs-AlGaAs bilayers, $a_{0}^{*} = 12.03$ nm and Ry$^{*} = 4.64$ meV corresponding
                     to the values $\varepsilon = 12.9$ (times the vacuum dielectric constant $\varepsilon_{0}$), $m_{\mathrm{e}} = 0.07$ and
                     $m_{\mathrm{h}} = 0.30$ (in units of the electron mass).


\bibitem{Natascia-99} N. Andrenacci {\em et al.}, Phys. Rev. B {\bf 60}, 12410 (1999).

\bibitem{Natascia-03} N. Andrenacci, P. Pieri, and G.C. Strinati, Phys. Rev. B {\bf 68}, 144507 (2003).

\bibitem{Yip-03} S.T. Wu and S. Yip, Phys. Rev. A {\bf 67}, 053603 (2003).

\bibitem{cinesi-06} L. He, M. Jin, and P. Zhuang, Phys. Rev. B {\bf 73}, 214527 (2006), and references therein.

\bibitem{SadeMelo-06} A similar behavior was recently found for ultracold Fermi atoms by
 M. Iskin and C.A.R. S\'{a} de Melo, Phys. Rev. Lett. {\bf 97}, 100404 (2006), and cond-mat/0606624.

\bibitem{Bedaque-03} P.F. Bedaque, H. Caldas, and G. Rupak, Phys. Rev. Lett. {\bf 91}, 247002 (2003).

\end{thebibliography}
\end{document}